# INVESTIGATION OF BREAKDOWN INDUCED SURFACE DAMAGE ON 805 MHZ PILLBOX CAVITY INTERIOR SURFACES

M. R. Jana, M. Chung, M. Leonova, A. Moretti, A. Tollestrup and K. Yonehara,
Fermi National Accelerator Laboratory, Batavia, IL - 60510, USA
B. Freemire, and Y. Torun, IIT, Chicago, IL - 60616, USA
D. Bowring, LBNL, Berkeley, California - 94720, USA
G. Flanagan, Muons, Inc., Batavia, IL - 60134

*Abstract*
The MuCool Test Area (MTA) at Fermilab is a facility to develop the technology required for ionization cooling for a future Muon Collider and/or Neutrino Factory. As part of this research program, we have tested two 805 MHz vacuum RF cavities in a multi-Tesla magnetic field to study the effects of the static magnetic field on the cavity operation. This study gives useful information on field emitters in the cavity, dark current, surface conditioning, breakdown mechanisms and material properties of the cavity. All these factors determine the maximum accelerating gradient in the cavity. This paper discusses the image processing technique for quantitative estimation of spark damage spot distribution on cavity interior surfaces. The distribution is compared with the electric field distribution predicted by a computer code calculation. The local spark density is proportional to probability of surface breakdown and shows a power law dependence on the maximum electric field (E). This E dependence is consistent with the dark current calculated from the Fowler-Nordheim equation.

## INTRODUCTION

High intensity, low emittance muon beams are essential requirements for a future Muon Collider and/or Neutrino Factory. Low emittance muon beams can be produced by ionization cooling. This consists of passing muon beams through low-Z absorber material (liquid H), to reduce all components of the momentum and replacing only longitudinal momentum with accelerating fields using RF cavities. At the same time, to keep the muon beam focused, both the absorbing material and RF cavities are placed inside a strong magnetic field provided by a superconducting solenoid. The ionization cooling process is most efficient if both the accelerating fields and magnetic fields are high. To study the interactions of a static magnetic field with operation at high accelerating fields, two 805 MHz RF cavities e.g. (i) LBNL-Pillbox (ii) All-Season have been tested in a multi-Tesla magnetic field at the Fermilab MuCool Test Area (MTA). During operation with high magnetic field B = 2 - 5 T, the cavity's accelerating field degrades significantly, dark current and X-rays are produced and interior surfaces suffer severe breakdown damage [1-3]. In this paper, we present the spatial distribution of the breakdown damage spots on the interior surface of the RF cavity using ImageJ [4] and Mathematica software [5]. The measured number of breakdown spots per unit area is compared with both the electric field distribution (calculated by computer code) and the dark current calculated from the Fowler-Nordheim equation. Finally a power law dependence on the maximum electric field is discussed.

## 805 MHZ CAVITIES

*LBNL-Pill Box cavity*

A cross sectional view of the 805 MHz cylindrical pillbox cavity of inner diameter 312.4 mm is shown in Fig. 1. Two irises are covered by specially designed copper windows (separately shown). Each window has a radius of 97.9 mm. RF power at 805 MHz is fed in through a kidney shaped coupling slot on the cavity wall using WR975 rectangular waveguide [3] with 20 µs pulse duration and repetition rate of 15 Hz. The cavity is mounted inside a 44 cm bore diameter superconducting solenoid magnet which can generate magnetic fields up to 5 T. The peak electric field was 40 MV/m with RF power 4.2 MW at B=0 field and 16 MV/m with RF power 0.67 MW at B = 3 T.

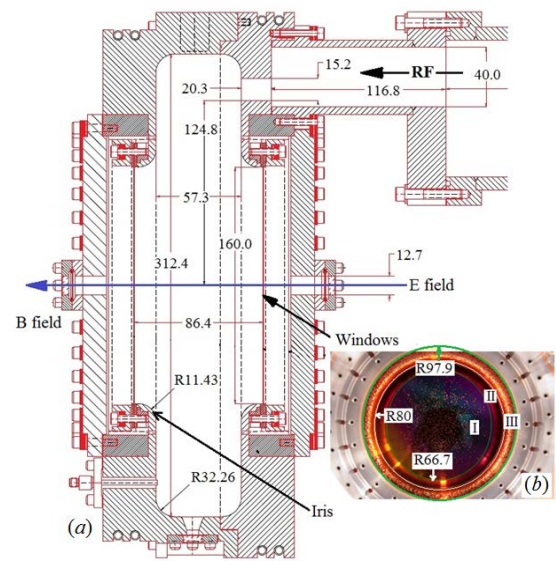

Figure 1: (*a*) Cross sectional view 805 MHz Pillbox RF cavity with copper window [3], (*b*) Copper window of radius 80 mm and iris of outer radius 97.95 mm.

Breakdown damaged spots on vacuum window is shown in Fig. 1(b) and the location of damaged spots are marked by crosshairs using ImageJ software shown in Fig. 2.

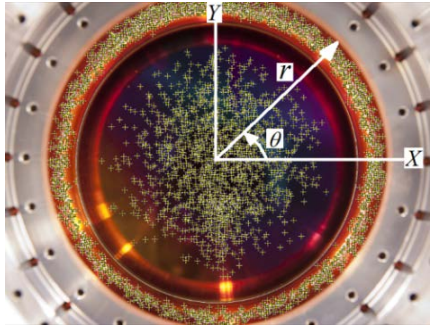

Figure 2: Photograph of copper window and iris with distribution of RF breakdown damage spots, marked with crosshairs using ImageJ software

*All Season Cavity*
Another cavity tested at the MTA known as the All Seasons cavity, is shown in Fig.3 [6]. This cavity was designed for both vacuum ($3\times10^{-8}$ Torr) and 100 atm compressed hydrogen gas operation. It consists of a SS316 cylindrical body [Fig. 3(b)] of outer diameter 366.9 mm and two end plates. The top [Fig.3(c)] and bottom [Fig. 3(d)] both have a diameter of 290.7 mm.

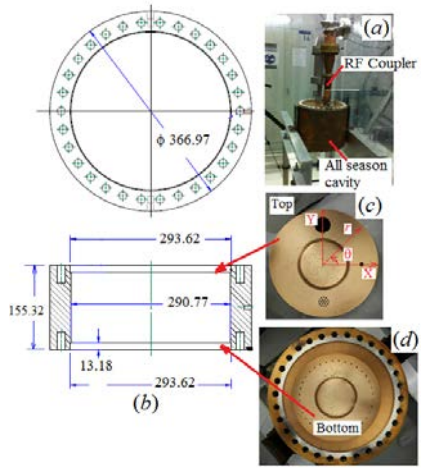

Figure 3: (a) All Seasons cavity (b) cross sectional view (c) top plate (d) bottom plate with cylindrical wall

RF power was fed to the cavity through a coax coupler [Fig. 3(a)] with 20 μs pulse duration and 15 Hz repetition rate. The cavity was operated at peak electric field of 25 MV/m with RF power 1.2 MW, B=0 T and 22 MV/m with RF power 0.93 MW, B= 3T magnetic field [7].

## ANALYSIS AND RESULTS

*LBNL-Pillbox cavity*
For the sake of analysis, the surface of the pillbox cavity window was divided into 3 regions as shown in Fig. 1(b). Region I is a circular area of radius 66.7 mm, region II is a ring of inner radius 66.7 mm and outer radius 80 mm. Both region I and region II are part of flat circular window. Region III (iris) is also a ring of inner radius 80 mm and outer radius 97.9 mm. The center of the window is the origin of the coordinate system (X, Y or $r,\theta$) shown in Fig.2 where Breakdown Damage (BD) spot locations are identified by crosshairs using the image processing software ImageJ. The BD spot distribution is calculated with the help of Mathematica software. Figure 4(*a*) shows the BD spot distribution in the X-Y plane. The upper half of the iris is close to the RF coupler and has more damage spots than the lower half. The BD spot distribution in r-θ plane is shown in Fig. 4(*b*).

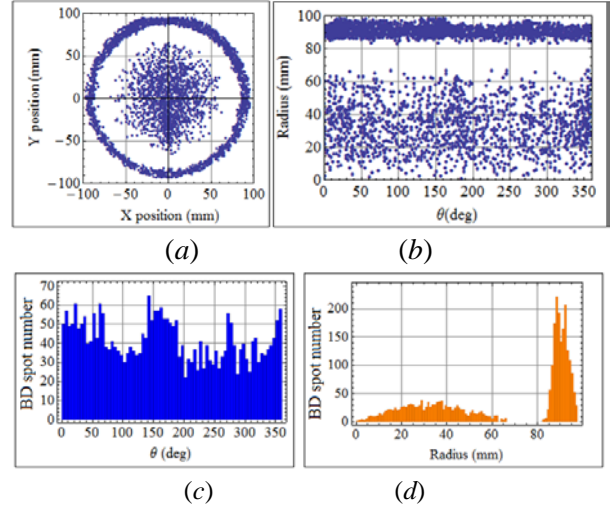

Figure 4: BD spot distribution in X-Y plane (*a*) and r-θ plane (*b*), BD spot no. vs θ plot (*c*) with bin size: 5° and BD spot no. vs *r* plot (*d*) with bin size 0.9 mm.

The number of BD spots at different values of θ and *r* are depicted in Fig. 4(*c*) and Fig. 4(*d*) respectively. The electric field distribution was calculated using computer code [8] and field emission dark current density was calculated from the Fowler-Nordheim equation where we have taken the work function (ϕ) for copper to be 4.65 eV and the field enhancement factor (β) to be 200. Figure 5 illustrates the results of the electric field, dark current density and number of BD spot density. Figure 6 shows the BD spot density vs electric field in the window area. The solid circles represent the data obtained from image processing and the line indicates an electric field ($E^n$) power fit. This shows the data fit well with $E^{6.28}$.

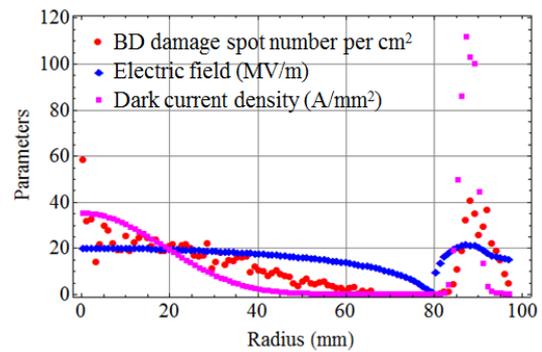

Figure 5: BD spot number per cm$^2$, RF electric field and dark current density vs radial distance plot

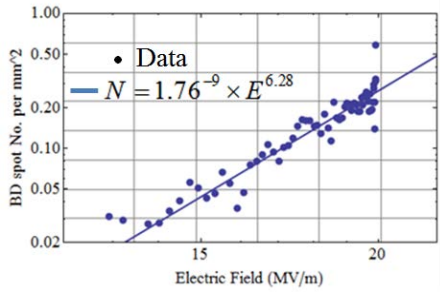

Figure 6: BD spot density vs electric field plot for flat window area.

*All Seasons Cavity*

Distributions of BD spot are shown in X-Y [Fig. 7(*a*)] and r-θ plane [Fig. 7(*b*)]. The number of BD spots at different values of θ and r are shown in Figs. 7(c) and 7(d) respectively. Fig. 8(a) illustrates the BD spot no. per mm$^2$ at different radii for the top plate at given θ values and Fig. 8(b) shows the same for the bottom plate. Fig. 9 depicts the computed electric field (with 10 times magnification) distribution and dark current density distribution. This is the same for both plates.

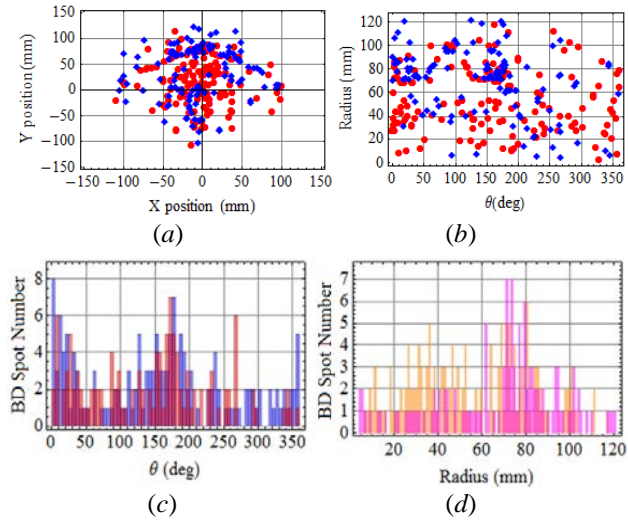

Figure 7: BD spot distribution in X-Y plane (*a*) and r-θ plane (*b*) (in both plots red/blue for top/bottom plate), BD spot no. vs θ plot (c) [blue/brown for top/bottom plate with bin size 5º], BD spot number vs *r* plot (d) [orange/magenta for top/bottom plate with bin size 1.2 mm]

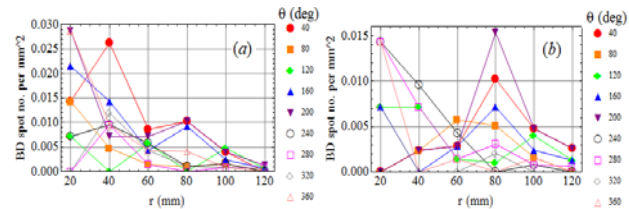

Figure 8: Plot of BD spot number per mm$^2$ at different radius for given θ (e.g. Red Circle for 40º, Orange Square for 80º and so on) (*a*) for top plate and (*b*) for bottom plate.

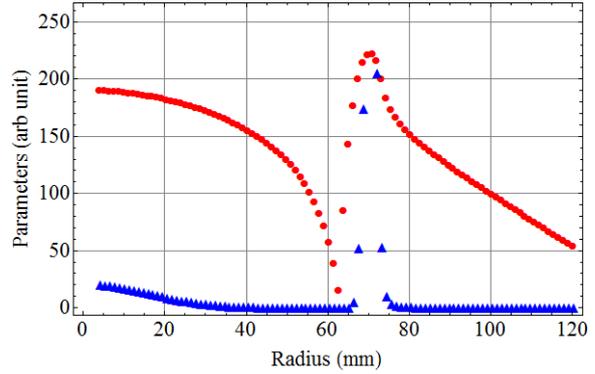

Figure 9: Electric field and dark current density distribution at different radii.

## CONCLUSIONS

The BD spot distributions in both cavities are generally consistent with the electric field profiles but there are some differences. In the LBNL-Pillbox cavity, BD spots were distributed more densely over the iris region where the electric field and dark current are high. Note that the iris region includes accumulated damage from more runs than the window. In the flat window region the BD spot density data correlates with 6.28$^{th}$ power of the electric field. In the All Season cavity, fewer damage spots are observed because the surface inspection was carried out at an early stage of the run. In both plates most of the spots are within a radius of 100 mm. Note that in both cavities, data was taken both with and without magnetic field before the surface inspection. In future experiments, inspection will be done separately after runs with and without magnetic field to explore the role of magnetic field in the breakdown process.